\documentstyle[12pt]{article}
\catcode`@=11
\def\seceqaa{\@addtoreset{equation}{section}
           \def\theequation{A\arabic{equation}}}
\catcode`@=12
\begin{document}

\title{Derivation of the Effective Pion-Nucleon Lagrangian within 
Heavy Baryon Chiral Perturbation Theory} 
\author{{A. Misra$^{(1)}$} \thanks{e-mail: aalok@iitk.ernet.in},
{D. S. Koltun$^{(2)}$} \thanks{e-mail: koltun@urhep.pas.rochester.edu}\\
(1) Indian Institute of Technology, Kanpur 208 016, UP, India,\\ 
(2) University of Rochester, Rochester, NY 14627, USA}
\maketitle
\vskip 0.5 true in

\begin{abstract}
We develop a method for constructing the Heavy Baryon Chiral Perturbation 
Theory (HBChPT) Lagrangian, to a given chiral order, within HBChPT. 
We work within SU(2) theory, with only the pion field interacting with the
nucleon. The main difficulties, which are solved, are to develop
techniques for implementing charge conjugation invariance, and for taking the
nucleon on shell, both within the nonrelativistic formalism. We obtain complete
lists of independent terms in ${\cal L}_{\rm HBChPT}$ through O($q^3$) for off-
shell nucleons. Then, eliminating equation-of-motion
(eom) terms at the relativistic and nonrelativistic level (both within HBChPT),
we obtain ${\cal L}_{\rm HBCHPT}$ for on-shell nucleons, through O$(q^3)$.
The extension of  the method (to obtain on-shell ${\cal L}_{\rm HBChPT}$ 
within HBChPT) to higher orders is also discussed. 
\end{abstract}



\clearpage 

\section{Introduction}

The theory of strong interactions, Quantum Chromodynamics (QCD), has no known 
exact solutions at low energies (referred to as the nonperturbative regime). 
Among the non-perturbative methods used, as an alternative, effective 
field theories (EFT) have been quite popular and
successful.  
The EFT that has now become quite popular is generically referred to as
Chiral Perturbation Theory (ChPT). ChPT was studied systematically first by
Gasser and Leutwyler \cite{gl} for purely mesonic systems. It was 
later extended to include baryons (nucleons) by Gasser, Sainio and Svarc 
\cite {gss} (referred to as Baryon ChPT [BChPT]). The nonrelativistic limit of 
BChPT was taken by Manohar and Jenkins, and the theory is referred
to as Heavy BChPT [HBChPT] \cite {jm}.

In the method given in the recent literature, the nonrelativistic Lagrangian 
of Heavy Baryon Chiral Perturbation Theory 
(${\cal L}_{\rm HBChPT}$) is obtained by a ${1\over{\rm m}}$-reduction
\cite {bkm} of the relativistic Lagrangian ${\cal L}_{\rm BChPT}$.
This first requires a complete representation of ${\cal L}_{\rm BChPT}$.
It would be more efficient if one could construct ${\cal L}_{\rm HBChPT}$ 
directly, by working entirely within the framework of the 
nonrelativistic (effective) field theory, without going through
the ${1\over{\rm m}}$-reduction (or Foldy-Wouthysen's transformation
\cite {kt}).

The main goal in this paper is to develop a method to construct
${\cal L}_{\rm HBChPT}$ directly within the framework of HBChPT for
processes involving a single off-shell nucleon and arbitrary number of pions. 
[For this paper, weak and electromagnetic interactions will not be 
included in the EFT.]
Such an off-shell Lagrangian can find applications in 
nuclear HBChPT as applied to e.g.,
pion double charge exchange scattering off a nuclear target (that involves
nucleons in a bound state).
Much of the process of implementing the requirements of chiral symmetry,
hermiticity, and Lorentz symmetries proceeds by methods similar to those
already in the literature. However, a new technique is required and developed, 
for implementing charge conjugation invariance within the nonrelativistic 
framework.

In this paper, the method developed will
be used for generating complete lists of independent terms in the Lagrangian,
of O$(q^2)$ and O$(q^3)$. The extension of the method to construction of 
O$(q^4)$ terms will be reported elsewhere. 
We first construct the lists of independent off-shell terms; 
then we use equation of motion techniques to generate the list of independent
on-shell terms. This requires a new method of embedding the effects of 
relativistic equations of motion in the HBChPT formalism.

In \cite{em}, EM have obtained  HBChPT on-shell  lists through O$(q^3)$
by ${1\over{\rm m}}$-reduction of ${\cal L}_{\rm BChPT}$. Our method
produces the same on-shell lists, as well as the full off-shell lists, but 
working entirely within HBChPT. Our results are shown to be identical to those
of EM for the on-shell lists. We also demonstrate how to extend the (on-shell) 
method to higher orders.

The basic meson and nucleon operators used for the construction of
${\cal L}_{\rm (H)BChPT}$, referred to as
``building blocks", are introduced in Section ${\bf 2}$, in which we also 
implement
chiral and Lorentz symmetries, and hermiticity. Our method for including
charge conjugation invariance is given in Section ${\bf 3}$. The following 
section
reduces the number of independent terms, using several algebraic identities.
The complete lists of off-shell terms of the type  $A^{(n)}$ 
(See (\ref{eq:ABC})), for chiral order $n=2,3$ are given in Section ${\bf 5}$.
In Section ${\bf 6}$,  a method is given for obtaining  the on-shell  
${\cal L}_{\rm HBChPT}$
within HBChPT; using it, the on-shell limit of the HBChPT terms  
up to O$(q^3)$ is taken. Comparison is made with EM's paper
\cite {em} in which they obtain the same on-shell terms, but by following the 
longer ${1\over{\rm m}}$-reduction formalism. Some concluding
remarks are given in Section ${\bf 7}$.

\section{Preliminaries}

The nonrelativistic Lagrangian
(${\cal L}_{\rm HBChPT}$) is constructed in terms of building blocks which 
are  of two kinds. The first kind are field operators constructed out of 
pion fields and derivatives, referred to as pion-field-dependent building 
blocks. There are four of  them: vector, axial-vector, scalar and 
pseudo-scalar. [Since we do not consider vector or 
axial-vector mesons (or resonances) and photons in this paper, there is no 
need to include a tensor field operator. For the latter, see
\cite {e2}.] The second kind are the baryon building blocks, which act only on
the baryon fields. There are three of them  which survive the nonrelativistic 
reduction of the five types of Dirac tensors:
a vector and an axial-vector, and the unit scalar ${\bf 1}$. 

We use the non-linear realization of chiral symmetry 
in which the nucleon isospinor transforms non-linearly (with respect to pions). 
To implement this, one introduces a field operator
$u\equiv\sqrt{U}$, where $U=e^{i\phi\over F_\pi}$. Here, 
$\phi\equiv\vec\pi\cdot\vec\tau$, where $\vec\pi$ is the pion field, 
$\vec\tau$ is the nucleon isospin operator,
and $F_\pi$ is  pion decay constant.

The following are 
the pion-field dependent building blocks that will be used for construction of 
${\cal L}_{\rm HBChPT}$, listed
by their Lorentz property. It is instructive to note that because these 
four building blocks are independent of the Dirac tensors, they are the 
same in BChPT and HBChPT.

\begin{eqnarray} 
\label{eq:bb}
& & {\rm vector}:
\ {\rm D}_{\mu} = {\partial}_{\mu} + {\Gamma}_{\mu},
\ {\rm where}\ {\Gamma}_{\mu}\equiv {1\over 2}[u^{\dagger},{\partial}_{\mu}u];
\nonumber\\
& & \nonumber\\
& & {\rm axial-vector}:
\ u_{\mu}\equiv i(u^{\dagger}{\partial}_{\mu}u - 
u{\partial}_{\mu}u^{\dagger});\nonumber\\
& & \nonumber\\
& & {\rm scalar}:
\ {\chi}_+\equiv u^{\dagger}\chi u^{\dagger}+ u{\chi}^{\dagger}u,\nonumber\\
& & \nonumber\\
& & {\rm pseudo-scalar}:
\ {\chi}_-\equiv u^\dagger\chi u^\dagger-u\chi^\dagger u,\nonumber\\
\end{eqnarray}
where $\chi\equiv\rm M_\pi^2$ for this paper.
\footnote{In general,
$\chi\equiv 2B(s+ip)$, where $s$ and $p$  are the external  scalar
and pseudoscalar fields respectively.
The constant $B$ is related to the quark condensate 
(See \cite{gl}).}
As regards the chiral orders of the four pion-field-dependent building blocks, 
using the facts that the field operators $U,u\equiv{\cal O}(1)$ and that 
$\partial_\mu\equiv{\cal O}(q)$, one sees:
\begin{equation} 
\label{eq:chObb}
u_\mu,\rm D_\mu\equiv{\cal O}(q),
\ \chi_\pm\equiv{\cal O}(q^2).
\end{equation}

All four of the pion-field dependent building blocks are constructed in  
such a way that they each transform homogeneously under the non-linear chiral
transformation K (defined via 
${\rm R}u{\rm K}^\dagger={\rm K}u{\rm L}^\dagger$, where R, L are
SU(2) matrices corresponding to independent chiral rotations of the
right (R)- and left (L)- handed components of the quark
field). 
Invariance under the non-linear realization of chiral symmetry imposes 
the constraint that ${\cal L}_{\rm (H)BChPT}$ can not contain $\rm D_\mu$
in the combination $\langle[\rm D_\mu,\ ]_+\rangle$, where 
$\langle\ \rangle$ denotes the trace on nucleon isospin. 
With this restriction, the Goldstone bosons are 
guaranteed to remain massless in the chiral limit, and the Ward identities 
are automatically satisfied by the Lorentz-invariant and anomaly-free 
generating functional \cite {ths}.

A bilinear ${\bar{\psi}}{\cal O}\psi$
in HBChPT is rewritten in terms of: 

\begin{eqnarray} & & 
\label{eq:Hh}
{\rm H}\equiv e^{i{\rm m}v\cdot x}{1\over2}(1+\rlap/v)\psi,
\nonumber\\
& & {\rm h}\equiv e^{i{\rm m}v\cdot x}{1\over2}(1-\rlap/v)\psi,
\end{eqnarray}
 referred to as 
the ``upper" and ``lower" components of $\psi$ respectively. 
(Here the parameter $v_\mu$ represents the nucleon velocity, and $m$ is the 
nucleon mass.)
The bilinear
then becomes: 
\begin{equation} 
\label{eq:ABC}
{\bar{\psi}}{\cal O}\psi 
= {\bar{\rm H}}A{\rm H}+{\bar{\rm h}}B{\rm H}+{\bar{\rm H}}
\gamma^0B^\dagger\gamma^0{\rm h} - {\bar{\rm h}}C{\rm h}.
\end{equation}

A systematic path integral derivation for ${\cal L}_{\rm HBChPT}$ based
on a paper by Mannel et al \cite {mnnl}, starting from 
${\cal L}_{\rm BChPT}$ was first given by Bernard et al \cite {bkm}. 
As shown by them, after integrating
out h from the generating
functional, one arrives at ${\cal L}_{\rm HBChPT}$ :

\begin{equation} 
\label{eq:lag}
{\cal L}_{\rm HBChPT} = {\bar{\rm H}}\biggl( A +
\gamma^0 B^\dagger\gamma^0 C^{-1}B\biggr)\rm H,
\end{equation}
an expression in the upper components only i.e. for non-relativistic nucleons.

The baryon building blocks act on the baryon fields; for terms in  
${\cal L}_{\rm HBChPT}$
of the form ${\bar{\rm H}}(...)\rm H$
they are defined in terms 
of the nucleon velocity $v_\mu$ and the Pauli-Lubanski spin operator 
${\rm S}_\mu\equiv{i\over2}\gamma^5\sigma^{\mu\nu}v_\nu$.
For such terms all five types of Dirac tensors can be reduced to
$v_\mu,\rm S_\nu$ (and ${\bf 1}$). 

Terms of the ${\cal L}_{\rm (H)BChPT}$ constructed from products of 
building blocks will automatically be chiral invariant. 
Symbolically, a term in ${\cal L}_{\rm HBChPT}$ 
can be written as just a product of the building blocks [(1) and $v_\mu,\rm S_
\nu$] to
various powers (omitting $\rm H,{\bar{\rm H}}$ as will be done in the
rest of the paper):
\begin{equation} 
\label{eq:prodbb}
{\rm D}_\alpha^m 
u_\beta^n\chi_+^p\chi_-^q v_\sigma^l\rm S_\kappa^r\equiv(m,n,p,q)
\equiv{\rm O}(q^{m+n+2p+2q}).
\end{equation}

The chiral order follows from (\ref{eq:chObb}); note that only pion-field 
dependent building
blocks contribute to the chiral order (if constructing
$B$-type terms, then one also uses $\gamma^5\equiv$ O$(q)$).
The isospin trace is not written because only the number of the building 
blocks is
important. 

In (\ref{eq:prodbb}), using the algebra  of $\rm S$'s,
one  can show that it is sufficient to consider
$r=0$ or 1 only (See \cite{ths}).

Now, we have considered O$(\phi^{2n})$ (even intrinsic parity)
and  O$(\phi^{2n+1})$ (odd intrinsic parity) terms separately as it 
helps in ensuring completeness. The following is  used: 
$u_\mu,\chi_-\equiv{\rm O}(\phi^{2n+1})$ and 
$\rm D_\mu,\chi_+\equiv{\rm O}(\phi^{2n}), n=0,1,2,....$ 
$(\phi\equiv$ mesons).
Hence, if one were to construct O$(q^3)$ counter terms 
then the allowed 4-tuples (m,n,p,q) are given in Table 1. 

We discuss how to exploit other symmetries of QCD, which are:
Lorentz invariance, isosopin symmetry, parity (including intrinsic),
hermiticity and  charge conjugation invariance (CCI).

The first four symmetries can be implemented directly within the 
nonrelativistic formalism, as follows below. However, this is not obvious for 
charge conjugation invariance, because 
charge conjugation transformation relates the positive - and negative - energy 
sectors of the fermions. In this paper, there are no anti-nucleons, and hence 
the negative-energy field ``h" (See (\ref{eq:Hh})) has been integrated out of 
the 
nonrelativistic field theory. The new work has been in developing a method to 
ensure charge conjugation invariance directly in the framework of HBChPT, and 
is discussed in the next section. 

(1) Lorentz invariance :
Lorentz scalars (and pseudo-scalars) are constructed by contracting
$v_\mu,\rm D_\mu
$ (vectors, denoted by ${\cal V}_\mu$), 
$\rm S_\mu, u_\mu
$ (axial-vectors, denoted by ${\cal A}_\mu$), 
$\chi_+
$ (scalar), $\chi_-
$ (pseudo-scalar), with 
$\epsilon^{\mu\nu\rho\lambda}$ and $g_{\nu\rho}$.
The discrete symmetry, parity, is used later to eliminate the pseudo-scalars.
One considers the $\epsilon^{\mu\nu\rho\lambda}$-dependent and independent terms
separately.

To ensure that all Lorentz indices have been contracted, the following 
condition has to be satisfied by the powers of the building blocks in 
(\ref{eq:prodbb}):
\begin{equation} 
\label{eq:phlor}
(-)^{m+n+l+r}=1.
\end{equation}

(2) Isospin symmetry: Exact isospin symmetry is assumed, i.e. $m_u\neq m_d$ 
effects are ignored. The baryonic field is taken to be an iso-doublet: 
$\biggl(^p_n\biggr)$, and the mesonic field operators are constructed 
out of the 
field $\phi\equiv\vec{\pi}\cdot\vec{\tau}$ (and $\partial_\mu$ 
for the covariant derivative $\rm D_\mu$), where
 $\vec\tau$ are the Pauli nucleon
isospin operators and $\vec\pi$ is the pion-isovector field. 
Terms in ${\cal L}_{\rm (H)BChPT}$ are then isospin invariant. 

(3) Parity : The parity operation is defined by 
$\vec x\rightarrow-\vec x , t\rightarrow t$ and $\phi\rightarrow-\phi$ 
(for pseudoscalar mesons), or in a covariant notation: 
$x_\mu\rightarrow x^\mu,\phi\rightarrow-\phi$.

One can show that $u_\mu, {\rm S}_\mu$ transform as pseudovectors,
$v_\mu, {\rm D}_\mu$ transform as vectors and 
$\epsilon^{\mu\nu\rho\lambda}$ transforms
as a pseudotensor. Hence, for terms of the form (6) including 
$n\ u_\mu$'s, $q\ \chi_-$'s, 
$r\ \rm S_\mu$'s and $j$ tensors
$\epsilon^{\mu\nu\rho\lambda}$ (with $j=0$ or 1),
a counting of the odd-parity terms such that
\begin{equation} 
\label{eq:phparity}
(-)^{n+q+r+j}=1
\end{equation}
makes the overall parity of the term even.

(4) Hermiticity : From the definitions of $\rm D_\mu, u_\nu, \chi_\pm$ in 
(\ref{eq:bb}),
one can see that $(\rm D_\mu^\dagger, \chi_-^\dagger)=-(\rm D_\mu, \chi_-)$ and
$(u_\mu^\dagger,\chi_+^\dagger)=(u_\mu,\chi_+)$. Now, if one uses a set of 
hermitian field operators : $(i\rm D_\mu, u_\nu, \chi_+,i\chi_-)$, then one 
can define how to construct hermitian (anti-)commutators for all four field 
operators uniquely. Let
\begin{equation} 
\label{eq:prescp}
i^P({\cal O}_1,{\cal O}_2)\ \equiv
\ [{\cal O}_1,{\cal O}_2]_+,\ {\rm or}\  i[{\cal O}_1,{\cal O}_2];
\end{equation} 
this is hermitian for hermitian building blocks
${\cal O}_{1,2}$, where $P$ counts the number of commutators.

With the above considerations, one can consider hermiticity directly at the 
nonrelativistic level. (The nucleon building blocks $v_\mu$ and S$_\mu$ are
hermitian.) But because we shall need to  compare the 
hermiticity property with charge conjugation invariance in Section ${\bf 3}$, 
in the relativistic formalism, we return here to that formalism. 
Consider the following expression in BChPT:
\begin{equation}
\label{eq:bil}
i^P{\bar{\psi}}\Gamma\cdot({\cal O}_1,({\cal O}_2,(...,({\cal O}_{n-1},
{\cal O}_n))....))\psi,
\end{equation}
where $\Gamma\equiv$ any one of the 5 types of Dirac tensors. 
For $\Gamma$ and ${\cal O}_i$, the phases
$h_{\Gamma}$ and $h_i(\equiv h_{{\cal O}_i})$ 
are defined via $\gamma^0\Gamma^\dagger\gamma^0=(-)^{h_\Gamma}\Gamma$ and 
${\cal O}_i^\dagger=(-)^{h_i}{\cal O}_i={\cal O}_i$.
Taking the hermitian conjugate of (\ref{eq:bil}), one gets:
\begin{eqnarray}
\label{eq:herm} 
 & & \biggl(i^P{\bar{\psi}}\Gamma\cdot({\cal O}_1,
({\cal O}_2,...({\cal O}_{n-1},{\cal O}_n))....))\psi\biggr)^\dagger\nonumber\\
& & = (-)^{h_\Gamma+P}i^P
{\bar{\psi}}\Gamma
\cdot((...({\cal O}_n,{\cal O}_{n-1}),{\cal O}_{n-2}),....),
{\cal O}_2),{\cal O}_1)\psi,\nonumber\\
& & =(-)^{h_\Gamma}i^P{\bar{\psi}}\Gamma
\cdot({\cal O}_1,({\cal O}_2,(.....,({\cal O}_{n-1},{\cal O}_n))...))\psi.
\end{eqnarray}

\section{Charge Conjugation Invariance}

Unlike the other symmetries, charge conjugation invariance requires that one 
work in the relativistic formalism, and consequently CCI is defined in 
terms of BChPT. We show that CCI can be indirectly imposed in HBChPT, by 
requiring the invariance first in BChPT and then following its consequences into
HBChPT. For that purpose, one needs to remember that $v_\mu,\rm S_\nu$ in 
HBChPT can always be obtained from $i[\rm D_\mu,\ ]_+,\ \gamma^5\gamma_\nu$ 
in BChPT.

Define the phases $c_{\Gamma}$ and $c_{{\cal O}_i}\equiv c_i$ via 
${\cal C}^{-1}\Gamma{\cal C}=(-)^{c_\Gamma}\Gamma^T$, and as we will see below, 
${\cal O}_i^c=(-)^{c_i}{\cal O}_i^T$, where $T$ is the transpose.
Charge conjugating (\ref{eq:bil}), one gets:
\begin{equation} 
\label{eq:cci}
\biggl(i^P{\bar{\psi}}\Gamma\cdot({\cal O}_1,(..({\cal O}_{n-1},
{\cal O}_n)..)\psi\biggr)^c=(-)^{c_\Gamma+c_{\cal O}+P}
i^P{\bar{\psi}}\Gamma\cdot({\cal O}_1,(..({\cal O}_{n-1},{\cal O}_n)..)\psi,
\end{equation}
where $c_{\cal O}\equiv\sum_i^n c_i$.

One may obtain the charge conjugation properties of the four field 
operators from the QCD Lagrangian in the presence of external fields,
assuming it to be charge conjugation invariant.
By doing so, 
one gets $i\rm D_\mu^c=-i\rm D_\mu^T, s^c=s^T, p^c=p^T$ ($s ,\ p$ 
are external scalar and pseudo-scalar fields). Now, as 
$\phi\equiv\vec\pi\cdot\vec\tau$
transforms the same way under CCI as $p$, 
this implies $\phi^c=\phi^T$, which in
turn implies $u^c=u^T$, and hence $u_\mu^c=u_\mu^T$. Finally using 
$(s^c,p^c)=(s^T,p^T)$, one gets $\chi_\pm^c=\chi_\pm^T$ 
(See (\ref{eq:bb}) and footnote 1).
One finds that $i\rm D_\mu$ is the only charge-conjugate-odd building block.

Now, to ensure that CCI and hermiticity are satisfied simultaneously, the 
following condition must obtain (comparing (\ref{eq:herm}) and (\ref{eq:cci})):
\begin{equation}
\label{eq:phhercci}
 (-)^{c_\Gamma+c_{\cal O}+P}=(-)^{h_\Gamma}=1.
\end{equation}
This equation remains unchanged if one considers expressions with 
traces.

Next, we show how this phase relation can be reexpressed in the nonrelativistic
formalism. Because $v_\mu$ can be obtained from a non-relativistic reduction of
either $[i\rm D_\mu,\ ]_+$ or $\gamma_\mu$, and $\rm S_\nu$ from the
nonrelativistic reduction of $\gamma^5\gamma_\nu$, it is in fact sufficient to 
consider only ${\bf 1},\gamma_\mu,\gamma^5\gamma_\mu$ of the five types of 
Dirac tensors. Further, one need not consider $\gamma_\mu$, because 
any $v_\mu$-dependent- nonrelativistic term obtained from the reduction of a 
$\gamma_\mu$ - dependent - relativistic term, can also be obtained by a 
corresponding relativistic term with the $\gamma_\mu$ replaced by 
$[i{\rm D}_\mu,\ ]_+$. (Note that $\gamma_\mu$ and $[i{\rm D}_\mu,\ ]_+$ have 
the same hermiticity and charge conjugation properties). 
Also, as the hermiticity 
and charge conjugation properties of ${\bf 1}$ and $\gamma^5\gamma_\mu$ are 
the same, $\rm S_\mu$ does not affect the phase in (\ref{eq:phhercci}). 
So, finally it is 
sufficient to consider only $\Gamma\equiv{\bf 1}$. 
Given that $(-)^{c_\Gamma}=(-)^{h_\Gamma}=1$, for $\Gamma\equiv{\bf 1}$, 
(\ref{eq:phhercci}) reduces to:
\begin{equation}
\label{eq:master}
 (-)^{c_{\cal O}+P}=1.
\end{equation}

We see from the discussion in the paragraph following (\ref{eq:cci}) 
that  $(-)^{c_{\cal O}}$ is  
equivalent to the phase factor coming from counting the number
of $i\rm D_\mu$'s in a given term in the BChPT Lagrangian. This in turn 
equals the phase factor coming from counting the number of $v_\mu$'s and $i\rm 
D_\mu$'s in the equivalent HBChPT term obtained after taking the 
non-relativistic limit. Also,
$(-)^P$ is equivalent to the number of commutators in the HBChPT term, which 
is the same as the number of commutators in the corresponding BChPT term.
[Note that the number of anticommutators can change in the reduction from
BChPT to HBChPT, since $[i\rm D_\mu,\ ]_+$ in the former can produce $v_\mu$
in the latter. However, $P$ remains unaffected.]
So, one thus arrives at the following rule for constructing HBChPT terms that 
are hermitian and charge conjugation invariant:

Only those HBChPT terms are allowed, which consist of 
$l\ v_\mu$'s, $m\ i\rm D_\nu$'s and $P[\ ,\ ]$'s, and which are 
made hermitian using the prescription of (\ref{eq:prescp}),  for which 
the following equation holds true:
\begin{equation} 
\label{eq:phvDP}
(-)^{l+m+P}=1.
\end{equation}
It should be noted that (\ref{eq:phvDP}) is an equation for HBChPT 
terms; no Dirac
phases remain.
Lorentz invariance and parity can be used for a further simplification. 
First, use (\ref{eq:phlor}) to rewrite (\ref{eq:phvDP}) as
\begin{equation}
\label{eq:phparlor}
(-1)^{n+r+P}=1
\end{equation}
Then, combining this with (\ref{eq:phparity}) 
yields the $phase\ rule$ for charge conjugation
invariant terms:
\begin{equation} 
\label{eq:phcci}
(-1)^{q+P+j}=1.
\end{equation}

This means that HBChPT terms consisting of $q\ i\chi_-$'s, $P[\ ,\ ]$'s 
and $j$ ($=$ 0 or 1) $\epsilon^{\mu\nu\rho\lambda}$'s,
that are Lorentz scalars and isoscalars of even parity, and are made 
hermitian using the prescription of (\ref{eq:prescp}),
for which equation (\ref{eq:phcci}) is valid, are the only terms allowed.
For application of (\ref{eq:phcci}), it is 
assumed one first writes down a complete list of hermitian Lorentz 
scalar-isoscalars satisfying chiral symmetry using the prescription of
(\ref{eq:prescp}). The phase rule (\ref{eq:phcci}) 
is then used to pick out those HBChPT terms whose 
BChPT counterparts are also charge conjugation invariant.

We give some examples to illustrate the phase rule (\ref{eq:phcci}) in Table 2.

\section{Reduction due to Algebraic Identities}

In this section we use a variety of algebraic identities to
reduce the number of independent terms in the Lagrangian, from
that obtained by use of the symmetry rules in the previous two sections.

\subsection{Elimination of traces}

For exact isospin symmetry, we may use several identities for traces
on the isospin of the nucleons to show that trace-dependent terms may be
eliminated from the Lagrangian, in favor of trace-independent terms.
We classify field operators as isoscalar or isovector using the standard
Pauli representation
\begin{equation}
\label{eq:Pauli}
{\cal O}_i={\cal  O}^0_i{\bf 1}+{\cal O}_i^a\tau^a
\end{equation} 
where
${\bf 1}\equiv 2\times 2$ unit matrix in the isospin space. 
Then ${\cal O}_i^a$ is the isoscalar component of ${\cal O}_i$ for $a=0$, 
and is the isovector component (of ${\cal O}_i$) for $a=1,2,3$.
For example, $\chi_+$ is isoscalar;   
$\chi_-$ and $u_{\mu}$ are isovectors. 
Then, if ${\cal O}$ is isoscalar or isovector,
\begin{equation}
\label{eq:trisosc}
\langle{\cal O}\rangle  =  2{\cal O},
\end{equation}
or
\begin{equation}
\label{eq:trisovec}
\langle{\cal O}\rangle  =  0.
\end{equation}
These relations hold for ${\cal O}$ which are functions of the basic field 
operators, as well. The one combination of field operators which cannot
appear within a trace is $[{\rm D}_\mu,{\cal O}_j]_+$, since that would
violate chiral symmetry (see Section ${\bf 2}$). 
In what follows, we exclude that 
operator from the ${\cal O}_i$.
For all other operators, the coefficients of ${\bf 1}$ or $\tau^a$ commute.
Thus, the trace 
\begin{equation}
\label{eq:zerotrace}
\langle[{\cal O}_i,{\cal O}_j]\rangle=0.
\end{equation}

Similarly, it is easily shown, using 
(\ref{eq:Pauli}) and the algebra of $\tau$-matrices, 
that if ${\cal O}_i$ and
${\cal O}_j$ are both isoscalar, or both isovector,
\begin{equation}
\label{eq:nonzerotr}
\langle[{\cal O}_i,{\cal O}_j]_+\rangle=2[{\cal O}_i,{\cal O}_j]_+.
\end{equation}
If one operator is isoscalar, and one isovector, then
\begin{equation}
\label{eq:zerotr}
\langle[{\cal O}_i,{\cal O}_j]_+\rangle=0.
\end{equation}

As a result of (\ref{eq:zerotrace}) and
 (\ref{eq:zerotr}), any trace-dependent term constructed
from basic field operators can be put into one of the  following four forms:
\begin{equation} 
\label{eq:4forms}
\langle[{\rm D}_\mu,{\cal O}_1]\rangle;\ \langle
[u_\nu,{\cal O}_2]_+\rangle;\ \langle[\chi_+,{\cal O}_3]_+\rangle;\ 
\langle[\chi_-,{\cal O}_4]_+\rangle.
\end{equation}

Now ${\rm D}_\mu={\partial}_\mu{\bf 1}+ \Gamma_\mu^a\tau^a$, where 
$\Gamma_\mu^a$
commutes with the other ${\cal O}_j$. It follows that for ${\cal O}_j$
isoscalar, that 
\begin{equation}
\label{eq:trDOnonzero}
\langle[{\rm D}_\mu,{\cal O}_j]\rangle=2[{\rm D}_\mu,{\cal O}_j],
\end{equation} 
while for ${\cal O}_j$ isovector
\begin{equation} 
\label{eq:trDOzero}
\langle[{\rm D}_\mu,{\cal O}_j]\rangle=0.
\end{equation}

We find that all the trace-dependent terms of either basic field operators,
or of the forms of (\ref{eq:zerotrace}) or (\ref{eq:4forms}), 
can be reduced to
equivalent forms without traces, using (\ref{eq:trisosc}), (\ref{eq:nonzerotr}) 
or (\ref{eq:trDOnonzero}), or else vanish, 
using (\ref{eq:trisovec}), (\ref{eq:zerotrace}), (\ref{eq:zerotr})
or (\ref{eq:trDOzero}). After this procedure, no trace-dependent
terms remain.
This result holds to all chiral orders.

\subsection{Algebraic Reduction For Trace-Independent Terms}

In this subsection, we introduce a number of algebraic identities which
lead to linear relations among terms constructed from ``building blocks,"
as in Sections  ${\bf 2}$ and ${\bf 3}$, and show how these may be used to 
reduce the number of independent terms of $O(q^3)$ in the Lagrangian. Since 
we can eliminate all trace-dependent terms, we discuss only trace-independent 
terms here.

First, the ``curvature relation" (which holds in the absence of external 
vector  and axial-vector fields) :
\begin{equation}
\label{eq:curvature}
[\rm D_\mu,\rm D_\nu]={1\over 4}[u_\mu,u_\nu],
\end{equation}
is  used extensively.

Second, we use the following three relations among commutators and/or
anticommutators.

(i) If $A,B$ are hermitian building-block field operators,
then :

\begin{equation}
\label{eq:ABA}
 [A,[B,A]] \equiv 2 ABA - [A^2,B]_+.
\end{equation}

Sometimes for  the purpose of comparison with EM's O$(q^3)$ list, one
elimnates $ABA$ and retains $[A,[B,A]], [A^2,B]_+$ ($A\equiv v\cdot u,
B\equiv v\cdot\rm D$ and $A\equiv u^\mu, B\equiv v\cdot\rm D$).

(ii) Jacobi identity :
\begin{equation}
\label{eq:Jacobi}
 [A,[B,C]] = [[C,A],B] + [[A,B],C]
\end{equation}

(iii) 
\begin{eqnarray}
\label{eq:abc}
 & & [A,[B,C]_+]_+ = ABC + ACB + {\rm h.c.}\nonumber\\
& & ABC + {\rm h.c.} = -[[B,C],A] + BCA + {\rm h.c.} 
\end{eqnarray}
So, out of $[A,[B,C]_+]_+, ABC + {\rm h.c.}, ACB + {\rm h.c.}, [[B,C],A]$,
one can take any two as linearly independent  terms, say $ACB + {\rm h.c.},
[[B,C],A] $.

The algebraic reductions associated with Levi-Civita (LC)
independent terms are summarized in Table 3.
(In Tables 3 and 4, the index $i$ is used to indicate the O$(q^3)$ terms 
which have been selected for the final lists given in Section 5, in 
tables 7 and 8.)

For terms with the Levi-Civita tensor, we use the following 
relations.

\begin{eqnarray} 
\label{eq:LCone}
& & \epsilon^{\mu\nu\rho\lambda}
{\rm S}_\lambda\biggl([{\cal A}_\nu, [{\cal B}_\mu,{\cal C}_\rho]_+]
\nonumber\\
& \equiv & \epsilon^{\mu\nu\rho\lambda}{\rm S}_\lambda\biggl([{\cal B}_\mu,
[{\cal A}_\nu,{\cal C}_\rho]]_++[[{\cal B}_\mu,{\cal A}_\nu],{\cal C}_\rho]_+
\biggr),
\end{eqnarray}
where ${\cal A},\ {\cal B},\ {\cal C}$, are given by: 
\begin{eqnarray}   
\label{eq:choices}
& (a) & u_\nu,\  {\rm D}_\mu,\ v_\rho v\cdot u,\nonumber\\ 
&  (b) & v_\nu v\cdot u,\ {\rm D}_\mu,\ u_\rho,\nonumber\\ 
& (c) & {\rm D}_\nu,\ u_\mu,\ v_\rho v\cdot u,\nonumber\\
& (d) & {\rm D}_\nu,\ u_\mu,\ u_\rho,\nonumber\\
& (e) & u_\nu,\ {\rm D}_\mu,\ u_\rho.
\end{eqnarray}

From case (d), one finds that 
\begin{equation}
\label{eq:LCtwo}
\epsilon^{\mu\nu\rho\lambda}{\rm S}_\lambda
[u_\mu,[{\rm D}_\nu,u_\rho]]_+=0,
\end{equation}
from which it follows
\begin{equation} 
\label{eq:LCthree}
\epsilon^{\mu\nu\rho\lambda}
{\bar{\rm H}}{\rm S}_\lambda u_\nu{\rm D}_\mu u_\rho{\rm H}+{\rm h.c.}
={1\over 2} \epsilon^{\mu\nu\rho\lambda}{\bar{\rm H}}{\rm S}_\lambda
[u_\nu,[{\rm D}_\mu,u_\rho]_+]\rm H.
\end{equation}

Finally, we replace (\ref{eq:LCone}) 
by the substitution ${\rm S}_\lambda\rightarrow 
v_\lambda$, with ${\cal A} _\nu\equiv{\rm D}_\nu,\ {\cal B}_\mu\equiv
u_\mu,\ {\cal C}_\rho\equiv{\rm D}_\rho$. Also:
\begin{eqnarray} 
\label{eq:LCfour}
& &  i\epsilon^{\mu\nu\rho\lambda}{\bar{\rm H}}v_\lambda
{\rm D}_\nu u_\mu{\rm D}_\rho{\rm H}+{\rm h.c.} \nonumber\\
&  & ={i\over 2}\epsilon^{\mu\nu\rho\lambda}
{\bar{\rm H}}v_\lambda([{\rm D}_\nu,[u_\mu,{\rm D}_\rho]_+]+
[{\rm D}_\nu,[u_\mu,{\rm D}_\rho]]_+){\rm H}.
\end{eqnarray}

The algebraic reductions  associated with LC-dependent terms are 
summarized in Table 4.

Thus, 
using (\ref{eq:curvature}) 
- (\ref{eq:LCfour}), one gets
a large reduction in the trace-independent terms of O$(q^3)$.

\section{The Lists of Independent Terms in Off-Shell ${\cal L}_{\rm HBChPT}$}

In this section we give complete lists of terms in the Lagrangian of  
O$(q^2)$, O$(q^3)$ for off-shell nucleons. 
What appears in the following lists are 
terms in the Lagrangian of type $A^{(n)}$ (See (\ref{eq:lag}) 
and \cite {bkm}), for $n=2,3$, 
that can be written as sums of independent terms 
\begin{equation}
\label{eq:expansion}
\sum_{n=2}^3\sum_i\alpha^{(n)}_i({\cal O}_i)^{(n)},
\end{equation}
where the low energy coupling constants (LECs) are given by the 
$\{\alpha^{(n)}_i\}$.
(Note that $\{\alpha^{(2)}_i\}\equiv\{``a_i"\}$ in \cite {em}.)

For each of the nonrelativistic terms of the types $A^{(2)}$ and $A^{(3)}$, 
we will give the relativistic counterparts as well, for two reasons.
First, comparison of the terms
shows that there is a one-to-one correspondence between the 
chiral orders of the nonrelativistic terms and their relativistic counterparts. 
This implies that the coupling constants of the   
nonrelativistic terms of O$(q^2)$ and O$(q^3)$
are not fixed relative to those of O$(q)$ and O$(q,q^2)$, respectively.  
(This further implies that reparameterization invariance (RI), which we discuss in
${\bf 7}$, (See \cite {lm}) imposes no constraints
for off-shell nucleons up to O$(q^3)$)
Second, in the next section,
we will consider the on-shell limit of the off-shell O$(q^2)$ and O$(q^3)$
terms (listed below in Table  5, 6, 7 and 8). The relativistic counterparts 
will help in this operation.
For the purpose of comparison with the lists of Ecker and Mojzis \cite {em} 
in Section ${\bf 6}$, the terms in 
these  tables are categorized into three types: type ${\cal A}$ corresponding to
the terms proportional to  the nonrelativistic  equation of motion;
type ${\cal B}$
 corresponding to those terms whose  realtivistic  counterparts can be
rewritten as linear combinations of relativistic terms one of which is 
proportional to the relativistic equation of motion; 
terms present in \cite {em} 's O$(q^2)$ list (labeled by a$_i$) 
and  O$(q^3)$ list (labeled by b$_i$).

\subsection{O$(q^2)$\ terms}

In this subsection, we list terms in $A^{(2)}$ (Tables  5 and 6)
separated into terms with $\phi^{2n}$ and $\phi^{2n+1}$,
i.e., even or odd powers of the  pion field  $\phi$.
(The same is done for $A^{(3)}$-type terms in the following subsection.)

One also gets some of the terms 
with fixed coefficients (i.e. independent of any LEC's
other than axial-vector coupling constant $g_A$) that are 
${1\over{\rm m^0}}$-suppressed from :
\begin{equation} 
\label{eq:crosstwo}
(\gamma^0B^\dagger\gamma^0C^{-1}B)^{(2)}.
\end{equation}
(See \cite {bkm}) 

\subsection{O$(q^3)$ terms}

The algebraic
reduction for trace-independent O$(q^3)$ terms
has been used to construct the tables 7 and 8.
One also gets O$(q^3)$ terms with O$(q^2)$ LEC's 
that are ${1\over{{\rm m^0}^{1,2}}}$-suppressed 
(analogous to (\ref{eq:crosstwo})) 
relative to $A^{(3)}$, from :
\begin{equation} 
\label{eq:crossthree}
\biggl(\gamma^0B^\dagger\gamma^0
C^{-1}B\biggr)^{(3)}.
\end{equation}
We will not write these  terms as we are interested in the number of 
independent O$(q^3)$ LEC's that are required at O$(q^3)$. It should be 
remembered that (\ref{eq:crosstwo}) and (\ref{eq:crossthree})  
$\in A^{(3)}$ for off-shell nucleons.

\section{On-Shell Reduction}

In  this section we show the effect of putting the nucleons on-shell 
(i.e. the external nucleons are free). We have developed a method to 
obtain a complete on-shell reduction working entirely within HBChPT.
The goal of this section is to give the rules within HBChPT for this on-shell 
reduction  up to  any chiral order and in particular 
of Tables  5, 6, 7 and 8 of ${\bf 5}$, for on-shell nucleons.

First in ${\bf 6.1}$, we derive a rule 
for carrying out on-shell reduction within HBChPT
for the $A^{(2)}$ and $A^{(3)}$-type 
terms obtained in Sections ${\bf 5.1, 5.2}$, 
and later for $A^{(n)}$ for all $n$. 
Next in ${\bf 6.2}$, we construct on-shell $B^{(n)}$ 
for all $n$,
to evaluate the on-shell $\gamma^0B^\dagger\gamma^0C^{-1}B$ 
(``cross terms") of (\ref{eq:lag}),  within
HBChPT. Specifically, we carry out the on-shell reduction
of the O$(q^3)$ cross terms.
This will reduce the number of independent terms;
the remaining terms so obtained (through O$(q^3)$) 
are labelled by $a_i$ in Table 6, and by $b_i$ in Tables 7 and 8.
These lists can be compared with 
similar lists given by Ecker and Mojzis (EM) \cite {em}, 
and can be shown to be completely equivalent.
However, to get a complete on-shell reduction, 
EM first had to go back to ${\cal L}_{\rm BChPT}$, and
then take its nonrelativistic limit. Our method works directly in HBChPT.
The match with EM's on-shell-reduced lists also serves as one 
check on the completeness of the off-shell lists (up to O$(q^3)$) obtained 
in Section ${\bf 5}$. 

\subsection{On-Shell Reduction of $A^{(n)}$}

Taking the on-shell limit is equivalent to eliminating equation-of-motion (eom) 
terms in the Lagrangian. 
At the nonrelativisitic level, the  $A$-type HBChPT terms obtained after 
${1\over{\rm m}}$-reduction of BChPT eom terms get eliminated 
by a different technique, which is discussed in this subsection.

EM \cite {em} have discussed
how to eliminate the nonrelativistic eom terms by suitable field redefinition 
of H, up to O$(q^3)$.
The ten terms of Tables 7 and 8 with $i=11,12,13,15,16,19,22,28,29,31$, 
are of this type, because 
they are explicitly proportional to the HBChPT eom (i.e. consist of 
$iv\cdot\rm D\rm H$). These terms are indicated by the label ${\cal A}$ 
in the last 
column, in Tables 7 and 8. We will not repeat the analysis of EM.
However, it is not sufficient to eliminate 
all nonrelativistic eom terms within the framework of HBChPT, to obtain a 
complete reduction for on-shell nucleons. As an example, it can be shown that 
a class of HBChPT terms
independent of  $iv\cdot\rm D\rm H$, but consisting of 
$\rm S\cdot\rm D\rm H$, can also be eliminated for on-shell nucleons
[e.g.$i=4$] (See (\ref{eq:SDO})). 

Krause \cite {k} used an interesting technique to get a reduction in the terms 
for on-shell baryons in SU(3) BChPT, that we shall modify to suit the SU(2)
case of this paper. We apply this method to 
the seven $A^{(3)}$-type terms with $i=4,9,10,14,24,26,27$ 
(indicated by ${\cal B}$ in Tables 7 and 8). 
[We believe our method to be different from that of EM \cite {em}, who have 
taken the SU(2) limit of Krause's SU(3) list to ensure completeness.] 
All seven terms are such that after expanding
the (anti-)commutators, they can be written as :
\begin{equation}
\label{eq:ODnr}
{\bar{\rm H}}{\cal O}_\mu\rm D^\mu\rm H+{\rm h.c.},
\end{equation}
which implies that for these terms 
the BChPT counterparts can be rewritten by ``introducing" a term 
$(i\rlap/\rm D-\rm m^0) \psi$, as will be shown below. 
This requires modifying Krause's technique to suit SU(2) (H)BChPT.

The basic idea is that a BChPT term written symbolically as:
\begin{equation} 
\label{eq:ODr}
{\bar{\psi}}(\Gamma{\cal O})_\mu{\rm D}^\mu\psi
\end{equation}
(with $\Gamma\equiv$ fundamental Dirac tensor, ${\cal O}\equiv$ product 
of building blocks, coupled to $(\Gamma{\cal O})_\mu$)
can be rewritten as:
\begin{eqnarray} 
\label{eq:intreom}
& & {\bar{\psi}}(\Gamma{\cal O})_\mu g^{\mu\nu}\rm D_\nu
\psi\equiv{\bar{\psi}}(\Gamma{\cal O})_\mu(\gamma^\mu\gamma^\nu
+i\sigma^{\mu\nu}
){\rm D}_\nu\psi\nonumber\\
& & \equiv -i{\bar{\psi}}(\Gamma{\cal O})_\mu\gamma^\mu(i\rlap/\rm D-\rm m^0)
\psi-i\rm m^0{\bar{\psi}}(\Gamma{\cal O})_\mu\gamma^\mu\psi\nonumber\\
& & +i{\bar{\psi}}(\Gamma{\cal O})_\mu\sigma^{\mu\nu}\rm D_\nu\psi.
\end{eqnarray}

Writing $\Gamma\sigma^{\mu\nu}=\sum_{i=1}^5a_i\Gamma_i$ (for $i\equiv$ S, PS, 
V, A, T) from the completeness of the $\Gamma_i$'s, 
except for $\Gamma\equiv{\bf 1},\ \gamma^5$,  
\footnote{For $\Gamma\equiv{\bf 1},\ \gamma^5$, the analysis is the same as
the one following (\ref{eq:lorsame}); the difference is that instead of 
(\ref{eq:lorsame}),  it is (\ref{eq:intreom}) that should
be referred to in the analysis. The choice 
$\Gamma\equiv\gamma^5$ is relevant for constructing on-shell $B$. 
(See ${\bf 6.2}$)}
there must be an $i=j$ 
such that the tensor  type of $\Gamma$ is the same as that of $\Gamma_j$.
There are then two possiblities: 
(a) the Lorentz indices of $\Gamma$ and $\Gamma_j$ are the same, or
(b) the Lorentz indices of $\Gamma$ and $\Gamma_j$ are different.

(a) Lorentz indices of $\Gamma$ and $\Gamma_j$ are the same:

Rewriting (\ref{eq:intreom}) one obtains:
\begin{eqnarray} 
\label{eq:lorsame}
(1-a_j){\bar{\psi}}(\Gamma{\cal O})_\mu{\rm D}^\mu\psi & = & 
-i{\bar{\psi}}(\Gamma{\cal O})_\mu\gamma^\mu(i\rlap/\rm D-\rm m^0)\psi - 
i{\rm m^0}{\bar{\psi}}(\Gamma{\cal O})_\mu\gamma^\mu\psi \nonumber\\ 
& & +\sum_{i\neq j}{\bar{\psi}}(a_i\Gamma_i{\cal O})_\mu\rm D^\mu\psi.
\end{eqnarray}
Since the first term on the RHS of (\ref{eq:lorsame}) 
is proportional to the BChPT eom,
it can be eliminated by suitable field redefinition of $\psi$. 
The second term (on the RHS of (\ref{eq:lorsame}))
is of a lower order than the original term, and hence can be discarded. 
The third term remains.
 
Now, as long as $a_j\neq 1$,  one sees 
that the ${\bar{\psi}}(\Gamma{\cal O})_\mu{\rm D}^\mu\psi$ can 
be rewritten as a linear combination of terms distinct from itself.
Thus, the nonrelativistic term obtained from the ${1\over{\rm m}}$-reduction 
of LHS of (\ref{eq:lorsame}) can be written as linear combination of 
other nonrelativistic terms obtained from ${1\over{\rm m}}$-reduction of
RHS of (\ref{eq:lorsame}) which are already present 
in the complete list of terms.
(We find $a_j\neq 1$ to the case for $i= 9, 14, 24, 26, 27$.)  
For the case $a_j=1$, (\ref{eq:lorsame}) 
implies that nonrelativistic terms  other than (\ref{eq:ODnr}),
obtained from the ${1\over{\rm m}}$-reduction of (\ref{eq:lorsame}), 
are going
to be related. As will be shown later, 
there is only one term for which one gets
$a_j=1$ in (\ref{eq:lorsame}): $i=21$.

(b) Lorentz indices of $\Gamma$ and $\Gamma_j$ are different (as is the case 
for $i=10$):

\begin{eqnarray} 
\label{eq:lordiff}
{\bar{\psi}}(\Gamma{\cal O})_\mu\rm D^\mu\psi  & = & 
-i{\bar{\psi}}(\Gamma{\cal O})_\mu\gamma^\mu
(i\rlap/\rm D-\rm m^0)\psi - i{\rm m^0}{\bar{\psi}}
(\Gamma{\cal O})_\mu\gamma^\mu\psi\nonumber\\ 
& & + \sum_i {\bar{\psi}}(a_i\Gamma_i{\cal O})_\mu\rm D_\nu\psi.
\end{eqnarray}
The implications are exactly the same as that of (\ref{eq:lorsame}) 
(for $a_j\neq 1$), 
except that the LHS of (\ref{eq:lordiff}) cannot be obtained
from the third term on the RHS of (\ref{eq:lordiff}).

The analysis of (\ref{eq:intreom}) gets simplified if the ${\cal O}_\mu$ in 
(\ref{eq:ODnr}) can be 
expressed as $\rm S_\mu{\cal O}({\cal O}\equiv$ building block).
Consider the term ${\bar{\rm H}}{\rm S}\cdot{\rm D}{\cal O}\rm H$+h.c. 
Its relativistic  
counterpart is ${\bar{\psi}}\gamma^5\rlap/{\rm D}{\cal O}\psi$+h.c., which 
can be rewritten as:
\begin{eqnarray} 
\label{eq:SDO}
 & & 
{\bar{\psi}}\gamma^5\rlap/{\rm D}{\cal O}\psi
+{\rm h.c.} \nonumber\\
& & \equiv i{\bar{\psi}}(-i\stackrel{\leftarrow}{\rlap/{\rm D}}-\rm m)
\gamma^5{\cal O}\psi + i{\rm m}{\bar{\psi}}\gamma^5{\cal O} \psi+{\rm h.c.}.
\end{eqnarray}
The first term can be eliminated by a redefinition
of $\psi$. The second term is not going to reduce to the 
required nonrelativistic term; hence ${\bar{\rm H}}{\rm S}\cdot{\rm D}
{\cal O}\rm H$+h.c. can be eliminated for  on-shell nucleons. 
(This eliminates the $i$ = 4 term.) 

Using (\ref{eq:intreom}) - (\ref{eq:SDO}), 
one arrives at a rule directly within the nonrelativistic
formalism for  elimination of those $A^{(n)},\ n=2, 3$ terms 
that are either of the type ${\bar{\rm H}}
{\rm S}\cdot{\rm D}{\cal O}\rm H+{\rm h.c.}$ 
or are such that their relativistic counterparts can be rewritten  as a linear
combination of other relativistic  terms one of which is the BChPT eom:
\begin{eqnarray} 
\label{eq:rule}
& & {\rm HBChPT\ terms\ of\ the\ type\ 
{\bar{\rm H}}{\rm S}\cdot{\rm D}{\cal O}\rm H\ +\ h.c.} \nonumber\\
& & {\rm or\ {\bar{\rm H}}{\cal O}^\mu\rm D_\mu\rm H\ +\ h.c.\ can\ be\
eliminated}\nonumber\\
& & {\rm except\ for}\ 
{\cal O}_\mu\equiv iu_\mu v\cdot u.
\end{eqnarray}
It is  understood that all (anti-)commutators  in the HBChPT Lagrangian
are to be expanded out until one hits the first ${\rm D}_\mu$, so that the
$A$-type HBChPT term can be put in the form 
${\bar{\rm H}}{\cal O}^\mu\rm D_\mu\rm H$
+h.c. 

The $A^{(2)}$-type terms $i=1, 2, 6, 7$, can be shown to be
eliminated for  on-shell nucleons by application of the abovementioned rule.
Then one finds agreement between the remaining terms in Tables 5 and 6 
with EM's list of $A^{(2)}$ terms, as indicated in the last column, where
$a_i$ labels the term as in EM.
Similarly, the $A^{(3)}$ terms already discussed can be eliminated for
on-shell nucleons, and are so denoted in Tables 7 and 8. The remaining
terms appear in the EM list, where $b_i$ indicates their terms.
With the addition of the exceptional term
$i$ = 21, discussed next, the agreement of our construction with that of 
EM  of $A^{(n)}$ through O$(q^3)$ is complete. 

The reason why ${\cal O}_\mu=iu_\mu v\cdot u$ is an exception to (\ref{eq:rule}) 
has to do with the fact that $\Gamma_\mu=\gamma_\mu$ in (\ref{eq:ODr}), 
as explained
below. The relativistic counterpart of the 
$i=21$ term: $i{\bar{\rm H}}u^\mu v\cdot u\rm D_\mu\rm H+{\rm h.c.}$ is:
\begin{eqnarray}
\label{eq:exception2}
 & & 
{\bar{\psi}}(iu^\mu\rlap/u\rm D_\mu+{\rm h.c.})\nonumber\\
& & = {\bar{\psi}}u^\mu\rlap/u\gamma_\mu
(i\rlap/\rm D-\rm m^0)\psi+\rm m^0{\bar{\psi}}u^\mu\rlap/u\gamma_\mu\psi
-{\bar{\psi}}\gamma_\rho\sigma_{\mu\nu}u^\mu u^\rho\rm D^\nu\psi+{\rm h.c.}
\end{eqnarray}
After some algebra, one gets:
\begin{eqnarray} 
\label{eq:linrel}
& & 
{\bar{\psi}}\biggl( i[\rlap/u,u^\mu]{\rm D}_\mu + {1\over 2}
\epsilon^{\mu\nu\rho\lambda}
\gamma^5\gamma_\lambda[[u_\mu,u_\rho],{\rm D}_\nu]_+
\biggr)\psi\nonumber\\
& &  = {\bar{\psi}}\biggl({1\over 2}\sigma^{\mu\nu}[u_\mu,u_\nu]
(i\rlap/\rm D-\rm m^0)+{1\over 2}\rm m^0\sigma^{\mu\nu}[u_\mu,u_\nu]\biggr)
\psi.
\end{eqnarray}
Thus, $i=21$ term drops out from the above relativistic equation (and is
therefore not eliminated as an eom term), which is  
equivalent to setting $a_j=1$ in (\ref{eq:lorsame}). 
The second  and the first terms 
on the LHS of (\ref{eq:linrel}) 
(after ${1\over\rm m}$-reduction) give the $i=27$ term,
and a term $i{\bar{\rm H}}[[v\cdot u,u^\mu],\rm D_\mu]\rm H$ which can be 
eliminated for off-shell nucleons (See Table 3). 
(Further, by using (\ref{eq:rule}), the $i=27$ term 
can also be eliminated for on-shell nucleons.)

To generalise, one sees that if in (\ref{eq:ODr}), 
$\Gamma_\mu\equiv\gamma_\mu$, 
then by the application of (\ref{eq:intreom}), 
that relativistic term, and hence its 
nonrelativistic counterpart, will not be eliminated for on-shell nucleons. 
Now, $\gamma_\mu$  gives $v_\mu$ after ${1\over{\rm m}}$
-reduction, which can  enter the HBChPT Lagrangian only as $v\cdot{\rm D}$ 
or $v\cdot u$ or $\epsilon^{\mu\nu\rho\lambda}v_\rho$. A term with
$v\cdot{\rm D}$ can be rewritten as linear
combinations of terms one (or more) of which  has a $v\cdot{\rm D}\rm H$ 
that can be eliminated by field redefinition of H. 
The remaining terms in the linear combination would either have already 
been included, or can be considered  as separate terms which can be 
obtained independently. Now, $\epsilon^{\mu\nu\rho\lambda}v_\rho (v\cdot u)$
can be obtained from the nonrelativistic reduction of 
$\epsilon^{\mu\nu\rho\lambda}i[{\rm D}_\rho,\ ]_+\rlap/u$. Also, because
${\rm D}^2+{\rm m}^2$ after ${1\over{\rm m}}$-reduction gives
$({\rm D}^2-2i{\rm m}v\cdot{\rm D})$ 
and that $\epsilon^{\mu\nu\rho\lambda}v_\rho(v\cdot u)^{l_1}(l_1\geq 1)$
can be obtained from  the nonrelativistic reduction of 
$\epsilon^{\mu\nu\rho\lambda}
i[{\rm D}_\rho,\ ]_+\rlap/u(iu\cdot\rm D)^{l_1-1}$,
one sees that the following class
of nonrelativistic terms will also serve as the exceptions to the above rule 
if one is to extend it to higher orders (i.e. beyond third order):
\begin{equation} 
\label{eq:exception3}
{\cal O}_\mu\equiv\Biggl(i^{m_1+l_5+l_7+1},\ {\rm or}\
\epsilon^{\nu\lambda\kappa\rho}\times\Omega\Biggr)\times u_\mu\Lambda
\end{equation}
with $l_1\geq1$,  
$\Omega\equiv 1(i$) for $(-)^{m_1+l_5+l_7+1}=-1(1)$, 
or
\begin{equation}
\label{eq:exception7}
{\cal O}_\mu\equiv\Biggl(i^{m_1+l_5+l_7},\ {\rm or}\
\epsilon^{\nu\lambda\kappa\rho}\times\Omega^\prime\Biggr)
\times {\rm D}_\mu\Lambda,
\end{equation}
with $l_1\geq1$, $\Omega^\prime\equiv 1(i$) for 
$(-)^{m_1+l_5+l_7}=-1(1)$.
In (\ref{eq:exception3})
and (\ref{eq:exception7}),
\begin{equation}
\label{eq:exception8}
\Lambda\equiv\prod_{i=1}^{M_1}{\cal V}_{\nu_i}
\prod_{j=1}^{M_2} u_{\rho_j}
(v\cdot u)^{l_1}u^{2l_2}\chi_+^{l_3}\chi_-^{l_4}
([v\cdot{\rm D},\ ])^{l_5}
({\rm D}_\beta{\rm D}^\beta)^{l_6}(u_\alpha{\rm D}^\alpha)^{l_7},
\end{equation}
where ${\cal V}_{\nu_i}\equiv v_{\nu_i}\ \rm or\ \rm D_{\nu_i}$.
The number of ${\rm D}_{\nu_i}$s in (\ref{eq:exception8}) 
(and in (\ref{eq:exception4}), below) equals $m_1(\leq M_1)$.
Assuming that Lorentz invariance, isospin symmetry, parity and hermiticity
have been implemented, the choice of the factors
of $i$ in (\ref{eq:exception3}) - (\ref{eq:exception8}) 
(and (\ref{eq:exception4})) 
automatically incorporates the phase rule (\ref{eq:phcci}).
In (\ref{eq:exception3}) - (\ref{eq:exception8}) (and (\ref{eq:exception4})),
it is only the contractions of the building blocks that
has  been indicated.
\footnote{For L.C-(in)dependent terms  in 
(\ref{eq:exception3}) - (\ref{eq:exception8})
with $l_7(>)\geq 1$, the coupling constants will be
fixed  relative to those of lower order terms, which is a consequence of
reparameterization invariance. 
In this discussion however ($unlike$ ${\bf 5}$),
we are interested in the number of independent terms, and not the
number of independent coupling constants.}  
It is understood that $v\cdot\rm D$ always enters 
as a commutator with another building block.
(The $i$=21 term can be obtained from (\ref{eq:exception3}) 
as a particular case.)
On comparing (\ref{eq:exception3}) - (\ref{eq:exception8})
 with (\ref{eq:ODr}), $\Gamma_\mu\equiv\gamma_\mu$. 
Thus, just as for $i=21$ term, (\ref{eq:exception3}) - (\ref{eq:exception8})
will not be eliminated for on-shell
nucleons by the application of (\ref{eq:rule}). 

\subsection{On-Shell Reduction of $B^{(n)}$}

For on-shell nucleons, the ``cross terms" of (\ref{eq:lag}) are
not contained in $A$, and to carry out a complete on-shell reduction,
one has to use (\ref{eq:lag}); the phase rule (\ref{eq:phcci}) (used for
construction of $A$) will not suffice. As an example, it is possible that for
on-shell nucleons, one may be able to eliminate [as an eom term, the form]
${\rm S}\cdot{\rm D}{\cal O}$
that one would get from $A^{(n)}$-type terms, but one may still get the same
term from $\biggl(\gamma^0B^\dagger\gamma^0C^{-1}B\biggr)^{(n)}$(``cross terms")
in (5), where  $B^{(n-1)}=\gamma^5{\cal O}$ and $B^{(1)}=i\gamma^5{\rm S}\cdot
{\rm D}$.
This is because of the following. The abovementioned $B^{(n-1)}$ comes 
from the nonrelativistic reduction of ${\bar{\psi}}\gamma^5{\cal O}\psi$,
(and $B^{(1)}$ comes from the Dirac term ${\bar{\psi}}(i\rlap/\rm D -\rm m)
\psi$).
Because $\gamma^5$, after ${1\over{\rm m}}$-reduction contributes only
to $B$ and not to $A$, and because for on-shell nucleons, the cross  terms 
can not always also be obtained from the $A$-type terms, it thus becomes 
necessary to take
the on-shell limit of $B^{(n-1)}$-type terms (which contribute at O$(q^n)$) in
addition to  taking the on-shell limit of $A^{(n)}$.

Thus, in order to construct the cross terms within 
HBChPT, one should be able to construct on-shell $B$ within HBChPT.

Using arguments similar to those used in ${\bf 6.1}$, 
one sees that the following gives the on-shell $B$:
\begin{eqnarray} \label{eq:exception4}
& &  B({\rm on-shell}) \equiv\Gamma_B\times   
\Biggl(i^{m_1+l_5+l_7},\ {\rm or}\
\epsilon^{\nu\lambda\kappa\rho}\times\Omega^{\prime\prime}
\Biggr)\nonumber\\
& & \times
\prod_{i=1}^{M_1}\biggl(v_{\nu_i}\ {\rm or}\ [{\rm D}_{\nu_i},\ ]\biggr)
\prod_{j=1}^{M_2} u_{\rho_j}\nonumber\\ 
& & \times(v\cdot u)^{l_1}
u^{2l_2}\chi_+^{l_3}\chi_-^{l_4}([v\cdot{\rm D},\ ])^{l_5}
([{\rm D}_\beta,\ ])^{2l_6}([{\rm D}^\alpha,\ ]u_\alpha)^{l_7},
\nonumber\\
& &   
\end{eqnarray}
where
 $(-)^{M_2+l_1+l_4+l_7}=(1)-1$  for L.C.-(in)dependent  
(\ref{eq:exception4}) if $\Gamma_B\equiv\gamma^5{\rm S}_\mu,
\gamma^5v_{[\mu}{\rm S}_{\nu]}$,
and $(-)^{M_2+l_1+l_4+l_7}=(-1)1$ for L.C.-(in)dependent 
 (\ref{eq:exception4}) if  $\Gamma_B\equiv\gamma^5,\gamma^5v_\mu,$
$\epsilon^{\mu\omega\sigma\delta}\gamma^5v_\sigma{\rm S}_\delta$   
\footnote{$\epsilon^{\nu\lambda\kappa\rho}\times
(\Gamma_B\equiv
\epsilon^{\mu\omega\sigma\delta}\gamma^5v_\sigma{\rm S}_\delta)
\times(......)$ in (\ref{eq:exception4}) is to
be treated as  L.C-independent, because product of even number of
L.C.'s can be written as products of metric.}
 (for even  parity of $B$). 
\footnote{Use is made of that 
${\bar{\rm h}}\biggl(\gamma^5,\gamma^\mu,\gamma^5\gamma^\mu,\sigma^{\mu\nu},
\gamma^5\sigma^{\mu\nu}\biggr)\rm H$ gives ${\bar{\rm h}}\biggl(
\gamma^5,\gamma^5{\rm S}^\mu,
\gamma^5 v^\mu,$ $i\gamma^5v^{[\mu}{\rm S}^{\nu]},
\epsilon^{\mu\nu\rho\lambda}\gamma^5v_\rho{\rm S}_\lambda\biggr)\rm H$ 
(omitting factors of 2).}
Also, $\Omega^{\prime\prime}\equiv1(i)$ if $(-)^{m_1+l_5+l_7}=1(-1)$  for
L.C.-independent terms, and $\Omega^{\prime\prime}\equiv1(i)$ if
$(-)^{m_1+l_5+l_7}=-1(1)$  for L.C.-dependent terms. 
It is understood that
$v\cdot{\rm D}$, $\rm D_{\alpha,\beta,\nu_i}$ always enter as a commutator
in (\ref{eq:exception4}),
so  that $v\cdot{\rm D}$ and $\rm D_{\alpha,\beta,\nu_i}$
can not  act on H; the requirement
of $\rm D_{\alpha,\beta,\nu_i}$ entering only as a commutator
can be relaxed only for $\Gamma_B\equiv\gamma^5{\rm S}^\mu$
(See ${\bf 6.1}$ and footnote 4).  It is however, still possible in some cases 
 to rewrite (\ref{eq:exception4}) as:
\begin{equation}
\label{eq:exception5}
(\Gamma_B {{\cal O}^\prime})^\mu\rm D_\mu.
\end{equation}
Such terms  are to be excluded in constructing on-shell $B$.
Thus, one need not consider $i\gamma^5[{\rm D}_\mu,u^\mu]\in B^{(2)}$.

One can show  that the contribution of the on-shell limit of 
$B^{(2)}$, 
gives one additional term: 
$i=6:\ i{\bar{\rm H}}[{\rm S}\cdot{\rm D},\chi_-]\rm H$;  $\gamma^5\chi_-$ 
can be
obtained as a special case of (\ref{eq:exception4}). 
One should  thus note that for on-shell nucleons, the LEC of the $i=6$ term 
is that of the term $\gamma^5\chi_-\in B^{(2)}$. With the addition  of
$i=6$ term, the agreement of our construction with that of \cite{em}
of on-shell (\ref{eq:lag}) through O$(q^3)$ is  complete. 

So, terms of the form (\ref{eq:exception4})
appear in the on-shell list of $B$ unless they can be rewritten as
(\ref{eq:exception5}). Combining the conclusions of ${\bf 6.1}$
and ${\bf 6.2}$,
equations (\ref{eq:rule}), (\ref{eq:exception3}) - (\ref{eq:exception5}) 
[modulo algebraic reductions] give rules for the construction of
the Lagrangian  (\ref{eq:lag}) on-shell, within
HBChPT.

\section{Conclusion and Discussion}

The goal of this paper is to develop a method  
to generate a complete expansion of ${\cal L}_{\rm HBChPT}$ to a given
order directly within the framework of HBChPT.  
A new method has been 
developed to implement CCI (combined with other symmetries discussed 
in ${\bf 2}$) within the framework of HBChPT, resulting in
a phase rule derived in the paper. Additional  reduction of the number of
independent terms, using algebraic identities was subsequently carried out, 
which, incidentally, allows the omission of traces
(for  exact isospin symmetry). The method  was applied 
to generate complete lists of $A^{(2),(3)}$ for off-shell nucleons.
We obtain 8 O$(q^2)$ and 31 O$(q^3)$ independent terms, with undetermined 
low energy coupling constants (LECs).
The method has also been carried out to next order, for terms of 
O$(q^4,\phi^{2n})$, to be discussed 
in a future publication. 

The main advantage of  the method developed in the paper
for constructing the off-shell ${\cal L}_{\rm HBChPT}$, 
as compared to the standard 
${1\over{\rm m}}$-reduction formalism in the literature \cite {bkm}, 
\cite {em}, is the following. 
Unlike the standard
${1\over{\rm m}}$-reduction formalism, one is not required to know
(and therefore to construct) the exact form of
${\cal L}_{\rm BChPT}$, and then to perform the non-relativistic term-by-term 
reduction of ${\cal L}_{\rm BChPT}$, in order to
construct ${\cal L}_{\rm HBChPT}$. 
Thus, the phase rule method of this paper is more 
efficient than the standard ${1\over{\rm m}}$-reduction formalism 
in the literature, for off-shell nucleons.
In the present method, one-to-one 
correspondence between (the chiral orders of) ${\cal L}_{\rm HBChPT}$ and 
its counterpart ${\cal L}_{\rm BChPT}$, is automatically implied.
We have verified this in detail through O$(q^3)$, but there are
exceptions, e.g. in certain terms of O$(q^4,\phi^{2n})$, to be discussed 
in a future publication. 
The exceptions  arise due to reparameterization invariance (RI)
according to which the Lagrangian density is invariant under infinitesimal
variations of  the nucleon-velocity parameter. As
a consequence, the coupling constants of certain terms at a given chiral
order get fixed relative to those  of some terms at lower chiral orders. 

Also, a method is given for obtaining 
the on-shell terms of types $A$ and $B$, and hence 
the HBChPT  Lagrangian, entirely within HBChPT.  
For the purpose of comparison with  EM's complete lists of $A^{(2),(3)}$-type
terms for on-shell nucleons \cite{em}, the lists of $A^{(2),(3)}$-type terms 
obtained in Section ${\bf 5}$ were reduced within HBChPT, after elimination of 
`equation of motion' (eom) 
terms (both at the nonrelativistic and relativistic levels),
using a rule developed in Section ${\bf 6}$ for arbitrary chiral orders.
After performing the on-shell reduction up to 
O$(q^3)$, agreement was found with the lists given in the EM paper.
By developing a method of
constructing the on-shell $B$ (within HBChPT), the  issue of  imposing RI 
directly within HBChPT gets partially addressed
in the sense that the LECs of some O$(q^n)$ terms get fixed relative
to the LECs of some O$(q^{n-1})$ $B$-type terms. (E.g., 
the LECs of the nucleon kinetic energy 
and the O$(q^2)$-correction to the Yukawa term which
are $\rm D^2$ and $i[{\rm S}\cdot{\rm D},v\cdot u]_+$ respectively,
get fixed
relative to the Dirac and Yukawa terms respectively.)
However, the RI constraints on $A$-type terms in higher order than $O(q^3)$
will  still need to be worked out (see footnote 2); this is beyond the scope
of this paper. 

\section*{Acknowledgements}

We would like to thank N.Kaiser, Ulf-G.Meissner and  M.Luke for
their continuous help, in terms of clarifications and preprints.
This research was supported in part by the U.S. Department of Energy
under Grant No. DE-FG02-88ER40425 with the University of Rochester.

\clearpage

\begin{table}[htbp]
\centering
\caption{Allowed 4-tuples for O$(q^3)$ terms}
\begin{tabular} {|cc|cc|} 
\hline
\multicolumn{2}{|c|}{O$(q^3,\phi^{2n})$} 
& \multicolumn{2}{c|}{O$(q^3,\phi^{2n+1})$} \\ \hline
(3,0,0,0) & (1,2,0,0) & (0,3,0,0) & (2,1,0,0) \\
(1,0,1,0) & (0,1,0,1) & (0,1,1,0) & (1,0,0,1) \\ 
\hline
\end{tabular}
\end{table}

\begin{table}[htbp]
\centering
\caption{Examples of application of the phase rule (\protect\ref{eq:phcci})}
\begin{tabular} {|c|c|c|c|c|c|} 
\hline
${\rm HBChPT\ term}$ & $(-)^q$ & $(-)^P$ & $(-)^j$ 
& $(-)^{q+P+j}$ & ${\rm Y}\equiv{\rm allowed}$
\\ 
& & & & & ${\rm N}\equiv{\rm not\ allowed}$ \\ \hline
$\epsilon^{\mu\nu\rho\lambda} v_\mu i[i{\rm D}_\mu,[i{\rm D}_\rho,u_\lambda
]_+]$ & $+$ & $-$ & $-$ & $+$ & Y \\ \hline
$\epsilon^{\mu\nu\rho\lambda}\rm S_\mu[u_\nu,i[u_\rho,
i\rm D_\lambda]_+]_+$ & $+$ & $+$ & $-$ & $-$ & N \\ \hline
$[v\cdot u,i[i{\rm D}^\mu,u_\mu]]_+$ & $+$ & $-$ & $+$ & $-$ & N \\ \hline
$[{\rm S}\cdot u,i[i\chi_-,[i{\rm D}^\mu,u_\mu]_+]]_+$
& $-$ & $-$ & $+$ & $+$ & Y \\ 
\hline
\end{tabular}
\end{table}

\begin{table}[htbp]
\centering
\caption{Algebraic reduction in L-C-independent terms}
\begin{tabular} {|c|c|c|} \hline
Need not consider & If consider & Equations \\ \hline
$[u_\mu,[{\rm D}^\mu,{\rm S}\cdot{\rm D}]_+]_+;\ 
u\cdot{\rm D}{\rm S}\cdot{\rm D} +{\rm h.c.};$ & 
$i=1, 2, 4$ & (\ref{eq:curvature})-(\ref{eq:abc}) \\
$u_\mu{\rm S}\cdot{\rm D}{\rm D}^\mu+{\rm h.c.};\ 
[{\rm D}^\mu,{\rm S}\cdot{\rm D}],u_\mu];$ &
& \\
$[{\rm D}_\mu,[u^\mu,{\rm S}\cdot{\rm D}]_+]_+;\
{\rm D}\cdot u{\rm S}\cdot{\rm D}+{\rm h.c.}$; & & \\ 
$[[u^\mu,{\rm S}\cdot{\rm D}],{\rm D}_\mu];\
[[{\rm S}\cdot u,u^\mu],u_\mu]$; & & \\
$[[{\rm D}_\mu,u^\mu]_+,{\rm S}\cdot{\rm D}]_+$ & & \\ \hline
Same as above with 
$u_\mu\rightarrow v\cdot u;\ {\rm D}_\mu\rightarrow v\cdot{\rm D}$ 
& $i=7, 8, 12$ & (\ref{eq:curvature}) - (\ref{eq:abc})
\\ \hline
$[{\rm D}_\mu,[u^\mu,v\cdot u]_+]_+;\ i{\rm D}\cdot  u v\cdot u+{\rm h.c.}$;
& $i=21, 23, 29$ & (\ref{eq:curvature})-(\ref{eq:abc}) \\
$i[[u^\mu,v\cdot u],{\rm D}_\mu]:\ i[[v\cdot u,{\rm D}_\mu],u^\mu]$;
& & \\
$i[[{\rm D}_\mu,u^\mu],v\cdot u];\ iu\cdot\rm D v\cdot u +{\rm h.c.}$;
& &  \\
$i[v\cdot u,[u^\mu,\rm D_\mu]_+]_+;\ 
i[{\rm D}_\mu,[{\rm D}^\mu,v\cdot{\rm D}]]$;
& & \\ 
$i{\rm D}_\mu v\cdot{\rm D}{\rm D}^\mu$ & & \\ \hline
$i(v\cdot u)(v\cdot{\rm D})(v\cdot u)$ 
& $i=16, 17$ 
& (\ref{eq:ABA}) \\ \hline
$iu_\mu (v\cdot{\rm D})u^\mu$ &
$i=15, 18$ & (\ref{eq:ABA}) \\ \hline
\end{tabular}
\end{table}
\begin{table}
\centering
\caption{Algebraic reductions of L-C-dependent terms}
\begin{tabular} {|c|c|c|} \hline
Need not consider & If consider & Equations \\ \hline
$\epsilon^{\mu\nu\rho\lambda}v_\rho{\rm S}_\lambda\biggl([u_\nu,[{\rm D}_\mu
,v\cdot u]_+];
[v\cdot  u,[{\rm D}_\mu,u_\nu]_+]$; & $i=24, 25, 26$ & (\ref{eq:curvature}), 
(\ref{eq:LCone}), (\ref{eq:choices}) [(a)-(c)] \\
$[[{\rm D}_\mu,v\cdot u],u_\nu]_+;\ 
[{\rm D}_\mu,[{\rm D}_\nu,v\cdot{\rm D}]]_+;$ & & \\
$[{\rm D}_\mu,[u_\nu,v\cdot u]_+]\biggr)$ & & \\ \hline
$\epsilon^{\mu\nu\rho\lambda}{\rm S}_\lambda\biggl([u_\nu,
[{\rm D}_\mu,u_\rho]_+];\ u_\nu{\rm D}_\mu u_\rho$; & 
$i=27$ & (\ref{eq:curvature}), (\ref{eq:LCone}), (\ref{eq:choices})[(e)],\\
$[\rm D_\mu,[\rm D_\nu,\rm D_\rho]]_+\biggr)$ & & 
(\ref{eq:LCtwo}), (\ref{eq:LCthree})\\ \hline
$i\epsilon^{\mu\nu\rho\lambda}v_\lambda\biggl(
[{\rm D}_\nu,[u_\mu,{\rm D}_\rho]_+];\ 
[[u_\mu,{\rm D}_\nu],{\rm D}_\rho]_+;$ 
& $i = 3, 14$ & (\ref{eq:curvature}),  
(\ref{eq:LCone})[${\rm S}_\lambda\rightarrow v_\lambda$] \\
$[u_\mu,[\rm D_\nu,\rm D_\rho]]_+\biggr)$ & & (\ref{eq:LCfour}) \\ \hline
\end{tabular}
\end{table}

\begin{table} [htbp]
\centering
\caption{$A^{(2)}(\phi^{2n+1})$ terms}
\begin{tabular} {|c|c|c|c|} 
\hline
$i$ & $A^{(2)}(\phi^{2n+1})$ & ${\rm BChPT\ Counterparts}$ & Term \\ 
& & & type: ${\cal A},\ {\cal B}$; \\
& & &  a$_i$ in EM \\ \hline
1 & $i[{\rm S}\cdot{\rm D},v\cdot u]_+$ & 
${1\over{\rm m^0}}\gamma^5[\rlap/{\rm D},[u^\mu,\rm D_\mu]_+]_+$ & 
${\cal B}$ \\ \hline
2 & $i[{\rm S}\cdot u ,v\cdot{\rm D}]_+$ & 
${1\over{\rm m^0}}\gamma^5\Biggl[\rlap/u,
\Biggl({\rm D}^2+{\rm m}^2
-\biggl(-(i\rlap/{\rm D}-{\rm m}^0)^2$ & ${\cal A}$  \\
& & $+{i\over 8}\sigma^{\mu\nu}[u_\mu,u_\nu]\biggr)\Biggr)\Biggr]_+$ & \\ 
\hline
\end{tabular}
\end{table}
\begin{table} 
\centering
\caption{$A^{(2)}(\phi^{2n})$ terms}
\begin{tabular} {|c|c|c|c|} 
\hline
$i$ & $A^{(2)}(\phi^{2n})$ & ${\rm BChPT\ Counterparts}$ & Term \\ 
& & & type: ${\cal A},\ {\cal B}$; \\
& & & a$_i$ in EM \\ \hline
3 & $(v\cdot u)^2$ & ${1\over{{\rm m^0}^2}}[{\rm D}_\mu,u^\mu]_+^2$ & a$_2$ 
\\ \hline 
4 & $u^2$ & $u^2$ & a$_1$ \\ \hline
5 & $[{\rm S}_{\mu},{\rm S}_\nu][u^{\mu},u^{\nu}]$ 
& $i\sigma^{\mu\nu}[u_\mu,u_\nu]$ & a$_5$ \\ \hline
6 & $(v\cdot{\rm D})^2$ & 
${1\over{{\rm m^0}^2}}\Biggl({\rm D}^2+{\rm m}^2
-\biggl(-(i\rlap/{\rm D}-{\rm m}^0)^2$ & ${\cal A}$  \\
& & $+{i\over 8}\sigma^{\mu\nu}[u_\mu,u_\nu]\biggr)\Biggr)^2$ & \\ \hline
7 & ${\rm D}^2$ &  $\biggl(-(i\rlap/{\rm D}-{\rm m}^0)^2$ & ${\cal B}$ \\ 
& & $+{i\over 8}\sigma^{\mu\nu}[u_\mu,u_\nu]\biggr)$ &  \\ \hline
8 & $\chi_+$  & $\chi_+$ & a$_3$ \\ 
\hline 
\end{tabular}
\end{table}

\clearpage

\begin{table} [htbp]
\centering
\caption{$A^{(3)}(\phi^{2n+1})$ terms}
\begin{tabular} {|c|c|c|c|} 
\hline
$i$ & $A^{(3)}(\phi^{2n+1})$ & ${\rm BChPT\ Counterparts}$ & Term \\ 
& & & type: ${\cal A},\ {\cal B}$; \\
& & & b$_i$ in EM \\ \hline
1 & $[u^2,{\rm S}\cdot u]_+$ & $\gamma^5[u^2,\rlap/u]_+$ 
& b$_{11}$ \\ \hline
2 & $u^{\mu}{\rm S}\cdot u u_{\mu}$ & $\gamma^5u^\mu\rlap/u u_\mu$ 
& b$_{12}$ \\ \hline
3 & 
$i{\epsilon}^{\mu\nu\rho\lambda}
v_{\mu}[[u_{\nu},u_{\rho}],u_{\lambda}]_+ $
& 
$i{\epsilon}^{\mu\nu\rho\lambda}\gamma_{\mu}[[u_{\nu},u_{\rho}],
u_{\lambda}]_+$ & b$_5$ \\ \hline
4 & $[{\rm S}\cdot{\rm D},
[{\rm D}^{\mu},u_{\mu}]]$ & $\gamma^5[\rlap/{\rm D},
[{\rm D}_\mu,u^\mu]]$ & ${\cal B}$ \\ \hline
5 & $[{\rm S}\cdot u,{\chi}_+]_+$ & $[\gamma^5\rlap/u,\chi_+]_+$ 
& b$_{17,18}$ \\ \hline
6 & $i[{\rm S}\cdot{\rm D},{\chi}_-]$ & $i\gamma^5[\rlap/{\rm D},
\chi_-]$ & b$_{19}$ \\ \hline
7 & $v\cdot u {\rm S}\cdot u v\cdot u$ & 
${1\over{{\rm m^0}^2}}\gamma^5[{\rm D}_\mu,u^\mu]_+\rlap/u
[{\rm D}_\nu,u^\nu]_+$ & b$_{14}$ \\ \hline
8 & $[{\rm S}\cdot u ,(v\cdot u)^2]_+$ 
& ${1\over{{\rm m^0}^2}}\gamma^5[\rlap/u,[[{\rm D}_\mu,
u^\mu]_+,[{\rm D}_\nu,u^\nu]_+]_+]_+$ & b$_{13}$ \\ \hline
9 & $[{\rm S}\cdot u, {\rm D}^2]_+$ & $\gamma^5\biggl[\rlap/u, 
\biggl(-(i\rlap/{\rm D}-{\rm m}^0)^2$ & ${\cal B}$  \\
& & $+{i\over 8}
\sigma^{\mu\nu}[u_\mu,u_\nu]\biggr)\biggr]_+$ & \\ \hline
10 & ${\rm D}^{\mu}{\rm S}\cdot u{\rm D}_{\mu}$ & 
$\gamma^5{\rm D}_\mu\rlap/u{\rm D}^\mu-{1\over2}\gamma^5\Biggl[
\Biggl({\rm D}^2+{\rm m^0}^2$ & ${\cal B}$ \\
& & $-\biggl(-(i\rlap/{\rm D}-{\rm m}^0)^2+
{i\over 8}\sigma^{\mu\nu}[u_\mu,u_\nu]\biggr)\Biggr)$, &  \\
& & $\rlap/u\Biggr]_+$ & \\ 
\hline
\end{tabular}
\end{table}

\begin{table}[htbp]
\centering
\begin{tabular} {|c|c|c|c|} 
\hline
$i$ & $A^{(3)}(\phi^{2n+1})$ & ${\rm BChPT\ Counterparts}$ & Term \\ 
& & & type: ${\cal A},\ {\cal B}$; \\
& & & b$_i$ in EM \\ \hline
11 &  $[v\cdot{\rm D},[v\cdot{\rm D},{\rm S}\cdot u]]$ & 
${1\over{{\rm m^0}^2}}\gamma^5
\Biggl[\Biggl({\rm D}^2+{\rm m^0}^2 -\biggl(-(i\rlap/{\rm D}-{\rm m}^0)^2$
& ${\cal A}$ \\
& & $+{i\over 8}\sigma^{\mu\nu}[u_\mu,u_\nu]\biggr)\Biggr),$ & \\ 
& & $\Biggl[\Biggl({\rm D}^2
+{\rm m^0}^2-\biggl(-(i\rlap/{\rm D}-{\rm m}^0)^2
$ & \\
& & 
+ ${i\over 8}\sigma^{\mu\nu}[u_\mu,u_\nu]
\biggr)\Biggr),\rlap/u\Biggr]\Biggr]$ 
& \\ \hline
12 & $(v\cdot{\rm D}{\rm S}\cdot{\rm D}v\cdot u+{\rm h.c.})$ & 
${1\over{{\rm m^0}^2}}\Biggl({\rm D}^2+{\rm m^0}^2
-\biggl(-(i\rlap/{\rm D}-{\rm m}^0)^2$ & ${\cal A}$ \\
& & $+{i\over 8}\sigma^{\mu\nu}[u_\mu,u_\nu]\biggr)
\Biggr)\gamma^5\rlap/{\rm D}
[{\rm D}^\mu,u_\mu]_+$ &  \\ 
& & $+{\rm h.c.}$ & \\ \hline
13 & $[(v\cdot{\rm D})^2,{\rm S}\cdot u]_+$ & 
$\Biggl[\Biggl({\rm D}^2+{\rm m^0}^2
-\biggl(-(i\rlap/{\rm D}-{\rm m}^0)^2$ & ${\cal A}$ \\
& & $+{i\over 8}\sigma^{\mu\nu}[u_\mu,u_\nu]\biggr)\Biggr)^2,\gamma^5\rlap/u
\Biggr]_+$  & \\ \hline
14 & $i{\epsilon}^{\mu\nu\rho\lambda}v_{\mu}
{\rm D}_{\nu}u_{\rho}{\rm D}_{\lambda}$ & 
$i\epsilon^{\mu\nu\rho\lambda}\gamma_\mu{\rm D}_\nu u_\rho{\rm D}_\lambda$ 
& ${\cal B}$ \\ \hline
\end{tabular}
\end{table}

\clearpage

\begin{table} [htbp]
\centering
\caption{$A^{(3)}(\phi^{2n})$ terms}
\begin{tabular} {|c|c|c|c|} 
\hline
$i$ & $A^{(3)}(\phi^{2n})$ & ${\rm BChPT\ Counterparts}$ & Term \\ 
& & & type: ${\cal A},\ {\cal B}$; \\
& & & b$_i$ in EM \\ \hline
15 & $i[v\cdot{\rm D},u^2]_+$ 
& ${1\over{\rm m^0}}\Biggl[\Biggl({\rm D}^2+{\rm m^0}^2
-\biggl(-(i\rlap/{\rm D}-{\rm m}^0)^2$ & ${\cal A}$ \\ 
& & $+{i\over 8}\sigma^{\mu\nu}
[u_\mu,u_\nu]\biggr)\Biggr),u^2\Biggr]_+$ 
& \\ \hline
16 & $i[v\cdot{\rm D},(v\cdot u)^2]_+$ & 
${1\over{{\rm m^0}^2}}
\Biggl[\Biggl({\rm D}^2
+{\rm m^0}^2-\biggl(-(i\rlap/{\rm D}-{\rm m}^0)^2$
& ${\cal A}$ \\ 
& & $+{i\over 8}\sigma^{\mu\nu}[u_\mu,u_\nu]\biggr)\Biggr),
[{\rm D}^\mu,u_\mu]_+^2\Biggr]_+$ & \\ \hline
17 & $i[v\cdot u,[v\cdot{\rm D}, v\cdot u]]$ &
$ {1\over{{\rm m^0}^2}}\Biggl[[{\rm D}_\mu,u^\mu]_+,
\Biggl[\Biggl({\rm D}^2+{\rm m^0}^2$
& b$_3$ \\ 
& & $-\biggl(-(i\rlap/{\rm D}-{\rm m}^0)^2$ & \\
& & +${i\over 8}\sigma^{\mu\nu}[u_\mu,u_\nu]\biggr)
\Biggr),[{\rm D}_\nu,u^\nu]_+\Biggr]\Biggr]$ & \\ \hline
18 & ${1\over{\rm m^0}}[u^{\mu},[v\cdot{\rm D},u_\mu]]$ & 
${1\over{\rm m^0}}\biggl[u^\mu,
\Biggl[\Biggl([{\rm D}^2+{\rm m^0}^2-\biggl(-(i\rlap/{\rm D}
-{\rm m}^0)^2$ & b$_1$ \\ 
& & 
$+{i\over 8}\sigma^{\mu\nu}[u_\mu,u_\nu]
\biggr)\Biggr),u_\mu\Biggr]\Biggr]$ &
\\ \hline
19 & ${\epsilon}^{\mu\nu\rho\lambda}v_{\rho}{\rm S}_{\lambda}
[[u_{\mu},u_{\nu}],v\cdot{\rm D}]_+$ & 
${i\over{\rm m^0}}\sigma^{\mu\nu}\Biggl[[u_\mu,u_\nu],
\Biggl({\rm D}^2+{\rm m^0}^2$ & ${\cal A}$ \\
& & $-\biggl(-(i\rlap/{\rm D}-{\rm m}^0)^2$ &  \\ 
& & $+{i\over 8}\sigma^{\mu\nu}[u_\mu,u_\nu]\biggr)\Biggr)\Biggr]_+ $
& \\ \hline
\end{tabular}
\end{table}

\begin{table}[htbp]
\centering
\begin{tabular} {|c|c|c|c|} 
\hline
$i$ & $A^{(3)}(\phi^{2n})$ & ${\rm BChPT\ Counterparts}$ & Term \\ 
& & & type: ${\cal A},\ {\cal B}$; \\
& & & b$_i$ in EM \\ \hline
20 & 
${\epsilon}^{\mu\nu\rho\lambda}v_{\rho}{\rm S}_{\lambda}[u_{\mu},
[v\cdot{\rm D},u_{\nu}]]_+$ & $\sigma^{\mu\nu}\Biggl
[u^\mu,\Biggl[\biggl[{\rm D}^2+{\rm m^0}^2-\biggl(
-(i\rlap/{\rm D}-{\rm m}^0)^2$ & b$_{15}$ \\ 
& & $+{i\over 8}\sigma^{\mu\nu}[u_\mu,u_\nu]\biggr)
\biggr],u_\nu\Biggr]\Biggr]_+$ & \\ \hline
21 & $(iu^{\mu}v\cdot u{\rm D}_{\mu}+{\rm h.c.})$ & 
$(iu^\mu\rlap/u{\rm D}_\mu 
+ {1\over{\rm m^0}}(u\cdot\rm D)^2) + {\rm h.c.}$ 
& b$_4$ \\ 
\hline
22 & $i[u^2,v\cdot{\rm D}]_+$ & ${1\over{\rm m^0}}\Biggl[u^2,
\Biggl({\rm D}^2+{\rm m^0}^2
-\biggl(-(i\rlap/{\rm D}-{\rm m}^0)^2$ & ${\cal A}$ \\ 
& & $+{i\over 8}\sigma^{\mu\nu}[u_\mu,u_\nu]\biggr)\Biggr)\Biggr]_+$ 
& \\ \hline
23 & $i[u_\mu,[v\cdot u,\rm D^\mu]]$ & $i[u^\mu,[\rlap/u,\rm D_\mu]]$
& b$_2$ \\ \hline
24 & ${\epsilon}^{\mu\nu\rho\lambda}v_{\rho}
{\rm S}_{\lambda}[{\rm D}_{\mu},[u_{\nu},v\cdot u]]_+$ & 
${1\over{{\rm m^0}^2}}
\epsilon^{\mu\nu\rho\lambda}\gamma^5\gamma_\lambda[{\rm D}_\rho,
[u_\nu,[{\rm D}_\kappa,u^\kappa]_+]]_+,]_+$ & ${\cal B}$ \\ \hline
25 & ${\epsilon}^{\mu\nu\rho\lambda}v_{\rho}
{\rm S}_{\lambda}[u_\mu,[{\rm D}_\nu,v\cdot u]]_+$ &
${1\over{{\rm m^0}^2}}
{\epsilon}^{\mu\nu\rho\lambda}\gamma^5\gamma_\lambda
[{\rm D}_\rho,[u_\mu,[{\rm D}_\nu,[{\rm D}_\alpha, u^\alpha]_+]]_+]_+$ 
& b$_{16}$ \\ \hline  
26 & ${\epsilon}^{\mu\nu\rho\lambda}v_{\rho}{\rm S}_{\lambda}
[[{\rm D}_{\mu},u_{\nu}],v\cdot u]_+$ & 
${1\over{{\rm m^0}^2}}{\epsilon}^{\mu\nu\rho\lambda}\gamma^5\gamma_{\lambda}
[{\rm D}_\rho,[[{\rm D}_{\mu},u_{\nu}],[{\rm D}_\alpha,u^\alpha]_+]_+]_+$ 
& ${\cal B}$  \\ \hline
27 & $\epsilon^{\mu\nu\rho\lambda}{\rm S}_\rho
[\rm D_\mu,[u_\nu,u_\lambda]]_+$ &
$\epsilon^{\mu\nu\rho\lambda}\gamma^5\gamma_\rho
[\rm D_\mu,[u_\nu,u_\lambda]]_+$ & ${\cal B}$ \\ 
& & $+2i{\rm m}^0\sigma^{\mu\nu}[u_\mu,u_\nu]$ &  \\ \hline
28 & $i(v\cdot{\rm D})^3$ & ${1\over{{\rm m^0}^3}}
\Biggl({\rm D}^2+{\rm m^0}^2
-\biggl(-(i\rlap/{\rm D}-{\rm m}^0)^2$ & ${\cal A}$ \\ 
& & +
${i\over 8}\sigma^{\mu\nu}[u_\mu,u_\nu]\biggr)\Biggr)^3$ &  \\ \hline
\end{tabular}
\end{table}
\clearpage

\begin{table}[htbp]
\centering
\begin{tabular} {|c|c|c|c|} 
\hline
$i$ & $A^{(3)}(\phi^{2n})$ & ${\rm BChPT\ Counterparts}$ & Term \\ 
& & & type: ${\cal A},\ {\cal B}$; \\
& & & b$_i$ in EM \\ \hline
29 & $i[{\rm D}^2,v\cdot {\rm D}]_+$ & 
${1\over{\rm m^0}}
\Biggl[\Biggl({\rm D}^2
+{\rm m^0}^2-\biggl(-(i\rlap/{\rm D}-{\rm m}^0)^2$ 
& ${\cal A}$ \\ 
& & $+{i\over 8}\sigma^{\mu\nu}[u_\mu,u_\nu]\biggr)\Biggr),
\Biggl(-(i\rlap/{\rm D}-{\rm m}^0)^2$  & \\ 
& & $+{i\over 8}
\sigma^{\mu\nu}[u_\mu,u_\nu]\Biggr)\Biggr]_+$ & \\ \hline
30 & $[\chi_-,v\cdot u]$ & ${i\over{\rm m^0}}
[\chi_-,[{\rm D}_\mu,u^\mu]_+]$ & b$_6$ \\ \hline
31 & $i[v\cdot\rm D,\chi_+]_+$ & ${1\over{\rm m^0}}
\Biggl[\biggl(-(i\rlap/{\rm D}-{\rm m}^0)^2+
{i\over 8}\sigma^{\mu\nu}[u_\mu,u_\nu]\biggr),$ & ${\cal A}$
 \\
& & $\chi_+\Biggr]_+$ & \\ 
\hline
\end{tabular}
\end{table}


\begin{thebibliography}{99}


\bibitem{gl} J.Gasser and H.Leutwyler, Annals of  Physics 158
(1984) 142.
\bibitem{gss} J.Gasser, M.E.Sainio and A.Svarc, Nucl. Phys.B  307
(1988) 779.
\bibitem{jm} E.Jenkins and
A.V.Manohar, Phys. Lett. B  255 (1991) 558.
\bibitem{bkm} V.Bernard, N.Kaiser, J.Kambor and Ulf-G Meissner, Nucl.Phys.B
388 (1992) 315.
\bibitem{kt} J.G.Koerner J.G.Thompson, Phys. Lett. B  264 
(1991) 185. 
\bibitem{em} G.Ecker and M.Mojzis, Phys. Lett. B  365 (1996) 312.
\bibitem{e2} G.Ecker, J.Gasser, A.Pich and E. de Rafael, Nucl. Phys. B
321 (1989) 311.
\bibitem{ths} ${\rm A.Misra,\ Ph.D.\ Thesis}$, ${\it SU(2)\ Heavy\ Baryon\ 
Chiral\ Perturbation\ Theory\ for\ One-\ and}$ ${\it Two-\ Nucleon\ Processes;\
  Application\ to\ Pion\ Double\ Charge}$ ${\it Exchange\ to\ One\ Loop}$,
University of Rochester, 1997, Report UR 1508. 
\bibitem{mnnl} T.Mannel, W.Roberts. W.Ryzak, Nucl. Phys. B 368
(1992) 204.
\bibitem{lm} M.Luke and A.V.Manohar Phys. Lett.B 286 (1992) 348.
\bibitem{k} A.Krause, Helv. Phys. Acta 63 (1990) 3.
\end{thebibliography}
\end{document}